# Surface Folding Induced Attraction and Motion of Particles in a Soft Elastic Gel: Cooperative Effects of Surface Tension, Elasticity and Gravity

Aditi Chakrabarti and Manoj K. Chaudhury[*]

Department of Chemical Engineering
Lehigh University, Bethlehem, PA 18015

**ABSTRACT:** We report, for the first time, some experimental observations regarding a new type of long range interaction between rigid particles that prevails when they are suspended in an ultrasoft elastic gel. A denser particle submerges itself to a considerable depth inside the gel and becomes elasto-buoyant by balancing its weight against the elastic force exerted by the surrounding medium. By virtue of a large elasto-capillary length, the surface of the gel wraps around the particle and closes to create a line singularity connecting the particle to the free surface of the gel. Substantial amount of tensile strain is thus developed in the gel network parallel to the free surface that penetrates to a significant depth inside the gel. The field of this tensile strain is rather long range owing to a large gravito-elastic correlation length and strong enough to pull two submerged particles into contact. The particles move towards each other with an effective force following an inverse linear distance law. When more monomers or dimers of the particles are released inside the gel, they orient rather freely inside the capsules they are in, and attract each other to form close packed clusters. Eventually, these clusters themselves interact and coalesce. This is an emergent phenomenon in which the gravity, the capillarity and the elasticity work in tandem to create a long range interaction. We also present the results of a related experiment, in which a particle suspended inside a thickness graded gel moves accompanied by the continuous folding and the relaxation of the gel's surface.

*  e-mail: mkc4@lehigh.edu

# 1. INTRODUCTION

Interaction between particles mediated by the mechanical distortion of the surrounding medium has been the subject of considerable interests in physical[1,2], metallurgical[3] and biological[4-6] literatures. When atoms and solid particles are inserted in the bulk of a solid matrix[3,7], its elastic energy is increased that usually gives rise to a distance dependence repulsive force. Such forces are thought to play important roles in the dispersion of defects in an elastic medium. While elastic interactions of particles prevail in anisotropic fluids such as nematic liquid crystals[2], there are also examples[4-6] with proteins and other integral components of a cell exhibiting attractive as well as repulsive interactions via membrane mediated elastic forces. Interaction mediated by capillary and gravity forces[8-12] has also been the subject of considerable interests in the past and the present. In such cases, a large length scale emerges from the competition between gravity and capillarity that rules the range of interaction between particles dispersed on a liquid surface when the Bond number of the system is comparable to or greater than unity. Co-operative effects of surface tension and elasticity give rise to a plethora of other interesting phenomena[13-28]. However, to the best of our knowledge there has not been any report till this date regarding the co-operative roles of gravity, capillarity and elasticity in any type of attractive or repulsive interactions between particles. What we report here is a novel observation related to the long range interaction between rigid particles in a soft elastic gel, in which the surface of the gel folds to form singular line defects connecting the particles and the outer surface of the gel. These line singularities create a tensile strain field parallel to the surface that extends deep inside the gel and leads to some fascinating long range attractive interactions between the suspended particles.

When a particle is suspended in such a gel that has a gradient of thickness, it is found to move from the thicker to the thinner part of the gel due to the gradient of elastic strain energy.

## 2. EXPERIMENTAL SECTION

### 2.1 Materials

The spheres used for this study include two types of ceramic balls (fracture-resistant silicon nitride with density 3.25g/cm$^3$ and non-porous high alumina ceramic with density 3.9g/cm$^3$), copper balls (Alloy 102, 99.95% pure copper, density 8.94g/cm$^3$) and steel balls (bearing-quality E52100 alloy steel, hardened ball, density 7.8g/cm$^3$) that were purchased from McMaster-Carr. The diameters of the spheres used in the study range from 2mm to 6.4mm. The spheres were sonicated in acetone (General use HPLC-UV grade, Pharmco Aaper) in a Fisher Scientific Ultrasonic Cleaner (Model no. FS5) for 10 minutes after which they were blow dried with ultra-pure nitrogen gas. In some experiments, dimers were formed by joining two spheres with super glue (Scotch). The materials used in the preparation of the gel are N-(hydroxymethyl)-acrylamide (48% solution in water, Sigma Aldrich), potassium persulphate (99.99% trace metals basis, Sigma Aldrich), and N,N,N′,N′- tetramethylethylenediamine (TEMED, ≥99.5%, purified by redistillation, Sigma Aldrich). For most of the experiments a quartz cell (45mm x 30mm, 45mm high, Rame Hart p/n 100-07-50) was used to study the interaction between the spheres. Borosilicate Glass vials (27 mm diameter × 70 mm high) were purchased from Fisherbrand for use in the static Stokes experiment. These were cleaned with deionized (DI) water and blow-dried with nitrogen gas before use. All experiments were performed after placing the test cells on a 3d manipulated stage that was situated atop a vibration isolation table (Micro-g, TMC).

## 2.2 Preparation of Gel

A 3.1% (by weight of acrylamide monomer) physically cross-linked hydrogel was used for the present study. We provide below the salient features of the method used to prepare the gel based on what was described in our recent publication[28]. In a clean glass jar, N-(hydroxymethyl)-acrylamide (3.1 wt% basis) was added to DI water obtained from a Thermo Scientific Barnstead E-pure unit. This solution was stirred for 30 minutes with pure nitrogen gas bubble purging through it. This step was followed by the addition of 0.25% potassium persulphate and further stirring the solution for 20 minutes. The gelation begins within few minutes after the addition and stirring of the last ingredient, 0.3% TEMED, to the mixture. The gel solution was poured into the quartz cell and the glass vials immediately after all the ingredients were mixed to prepare the gel. The quartz cell was covered with Parafilm and the vials were tightly secured with their caps to avoid evaporation of water. The gelation was complete in 2 hours at room temperature. While the gels prepared as above were used for most of the experiments, in a couple of experiments (see figures 5 and 7) small amount (0.01 %) of N,N'- Methylene bisacrylamide (99%, Sigma Aldrich) was used as the crosslinking agent (see below) to increase its modulus slightly.

## 2.3 Elastic Stokes Experiment

This experiment was performed by gently placing either the ceramic, copper or steel spheres on the surface of the polyacrylamide (PAM) hydrogel cured in the vials, one at a time (Figure 2). The sphere immersed itself inside the gel and stood still at a depth of *h,* which was measured from the surface of the gel in the vial to the sphere's center. Care was taken to ensure that the spheres were at the centers of the vials in order to avoid putative wall effects. Even though some distortion of the shape of the spherical ball occurred when it was viewed through the sides of the cylindrical vial, no distortion of the image occurred in the vertical direction, which is what was needed for the measurement of *h*. The details of how the images were captured and processed are described in a later section. An experiment in which an alumina ceramic ball (3.2mm diameter) was released on the surface of the hydrogel prepared inside a rectangular quartz cell was captured by a high speed camera (Redlake Motion-Pro, Model no: 2000) at the rate of 500 frames per second. Snapshots from the high speed movie (Figure 1, A-D) gave an insight into the mechanism by which the gel surface wraps around the sphere, and relaxes with time.

**2.4 Interaction of Spheres inside the Gel**

A copper (2 mm or 2.4 mm) or an alumina ceramic (3.2 mm or 4.8 mm) sphere was released gently on the surface of the gel with the help of a prong holder (McMaster-Carr). The second sphere of same diameter as the first one was then released inside the gel in the same plane, perpendicular to the direction of the camera, within approximately 0.5- 1.0 cm away from the other sphere. The attraction and the descent of each pair of spheres in the gel were recorded with a CCD camera connected to a microscope for further analysis. In one experiment involving 4.8 mm alumina ceramic spheres, a dimer made of two glued balls (4.8 mm each) was released a distance of about 1 cm away from another dimer formed by the attractive contact of two other spheres in the same gel. The attraction of these dimers was also captured with a CCD camera. In the experiments illustrating the interactions of the clusters, the balls were released into the gel in such a way that two clusters grew by the self-assembly process not very far from each other. The interactions of the spheres, the growth of the clusters, their movements and attractions were all recorded with a CCD camera connected to a microscope as discussed below.

**2.5 Thickness Graded Gel**

A 3% chemically cross-linked PAM gel containing 0.01% (w/w basis) N,N'-Methylene bisacrylamide was used for the preparation of the thickness graded gel. The as-prepared gel solution was immediately poured into the Quartz cell that was then tightly secured with parafilm. In order to obtain a thickness gradient of the gel in the cell, it was inclined by elevating one end to a height of ~8mm. The gel was allowed to cure in the tilted cell for about 2 h after which the cell was brought back to its initial horizontal position. The gel surface was concave and flatter on both the edges due to the gel material sagging down from the edges, however there was a large length (~20mm; total length of the cell being 45mm) over which the gradient of thickness was more or less constant. The cell was placed on a 3d manipulated stage. A steel or a ceramic sphere was released on the thicker part of the gel. The video of the motion of the sphere down the gradient was captured using a CCD camera equipped with a variable focal length microscope.

**2.6 Videography and Analysis**

The interactions of the spheres were captured using a Video Microscope (Infinity) that was equipped with a CCD camera (jAi, Model no. CV−S3200) and connected to a computer using the WinTV application (Hauppauge, USA). The images for the Static Stokes experiment were also captured with the same video micrographic setup. The recorded videos were decomposed into image sequence in VirtualDub and the images were analyzed to measure the depth of submersion and the distance of separation between the spheres using a tracking algorithm, SpotTracker, in ImageJ. The calibration factor of the variable focal length microscope was obtained from the known diameter of the spheres in all the images.

**3. RESULTS AND DISCUSSIONS**

**3.1 Penetration of a single particle through the gel's surface** A millimeter size spherical object made of either ceramic, copper, or any metal submerges itself to a considerable depth[28] inside a soft hydrogel, the modulus of which is in the range of few Pascals. Even though the modulus of the gel is so low, its mesh size [$(k_B T/\mu)^{1/3} \sim 100$ nm] is still vanishingly small as compared to the size ($\sim$ mm) of the sedimenting object. The deformed network can exert sufficient elastic force to balance the weight of the sphere thus making it neutrally buoyant after it descends by a decent distance inside the gel. While the elastic field is symmetric around each sphere in the classic problems of elastic inclusions, the stress and the strain fields here are asymmetric (because gravity breaks the symmetry) in which the elastic stress beneath the sphere is higher than that above the sphere[28].

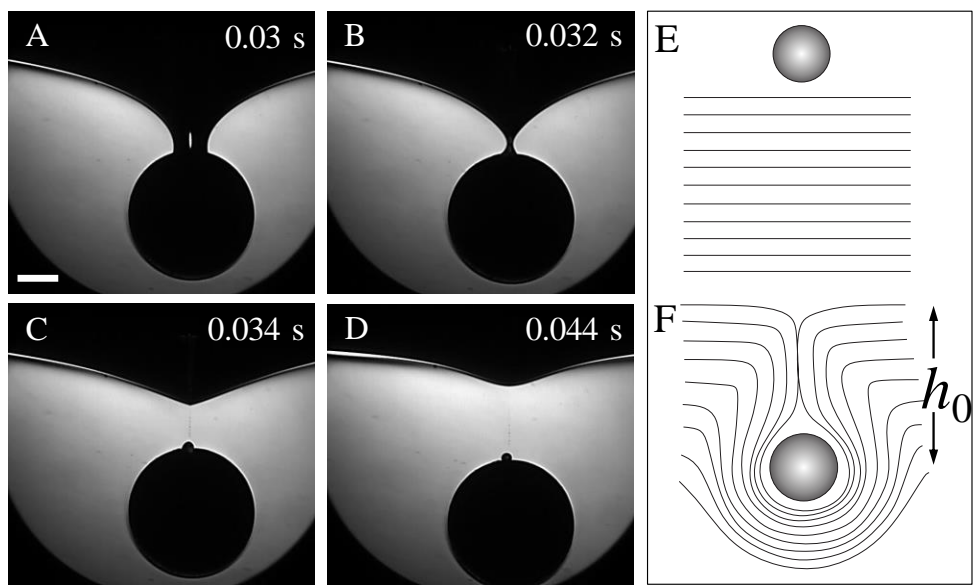

**Figure** 1. The fall of a small ceramic ball (3.2mm diameter) through a soft polyacrylamide hydrogel is captured with a high speed (500 frames per second) camera. The folding of the surface of the gel around the sphere (A), the pinch-off instability (B) and the formation of a thin line (C-D) connecting the ball and the surface of the gel are evident in these videographs. The surface of the gel relaxes slowly (D) with no sign of any fracture in the gel. The white scale bar here represents 1mm. (E-F) The schematic illustrates the wrapping of the sphere by and the folding of the surface of the gel as the sphere penetrates the gel. Here, $h_0$ is the initial height of a single ball immersed in the gel.

Figure 1 (A to D) depicts the case of a single ceramic ball released on the surface of a soft hydrogel, which sediments down to its equilibrium position within about 2 seconds. These high speed video micrographs do not provide any evidence of fracture in the gel. Instead the free surface of the gel wraps around the ball and folds into line contact above it. The contact region necks down so that a small bubble remains attached to the zenith of the sphere, while strings of very tiny air bubbles appear in the thin channel above it. These tiny air bubbles soon coalesce and escape through the narrow channel, which further closes due to auto-wetting forces of the gel's surface. This phenomenon has certain resemblance to the pinch-off instability[29,30] of a dripping liquid drop. However, the spherical ball here does not detach from the thin channel that connects (see Figure 3C and 3D) it to the free surface of the gel as the elastic force supports the weight of the ball[28]. A substantial stress concentration is expected to develop around the thin line joining the sphere and the free surface; thereby the region of the gel above the sphere remains in a state of tensile stress parallel to the surface (Figure 1F), the magnitude of which depends on the shear modulus, the radius of the sphere and how far the ball descends inside the gel. If the surface of the gel is pre-marked with ink spots, it is easy to visualize that the surface of the gel gets appreciably stretched while the sphere sinks through the gel while still connected to the free surface via thin channel. It is also possible to release ink inside the gel in the form of thin vertical lines with the help of a fine needle, which bend towards the sphere in a dramatic way when the sphere is released inside the gel[31]. The role of self-adhesion of the free surface of the gel above the sphere is crucial in keeping the folded surface intact against the tensile elastic field. It is surprising that the soft gel beneath the ball does not fracture, which we feel is a manifestation of two effects: blunting[32,33] of the crack brought about by the spherical geometry of the suspended object and the Lake-Thomas effect[34], in which the tearing energy of a rubber increases with the compliance of a network up to a certain extent. Intentional fracture can, however, be induced by the imposition of a stronger external field aided by thermal fluctuation, which we reserve for a detailed future study. In the absence of any disruptive external field, the sphere can be entirely supported by the elastic force, in which the depth of penetration increases with the mass ($m$) of the ball.

**3.2 Elastic Buoyancy.** The relationship between the depth of penetration ($h$) and the weight ($mg$) of the ball in the gel of shear modulus $\mu$ should be linear for small deformation, i.e. $h = mg/4\pi\mu R$ or, $h \sim R^2$ as was observed by us previously[28] with a higher modulus hydrogel that too underwent a large deformation although not as much as the current gel. This surprising observation suggested that a linear elastic modulus pleasingly scales out of the mechanics of the deformation of a system that is intrinsically neo-Hookean. However, this relation is non-linear for the lower modulus gel used in the current study.

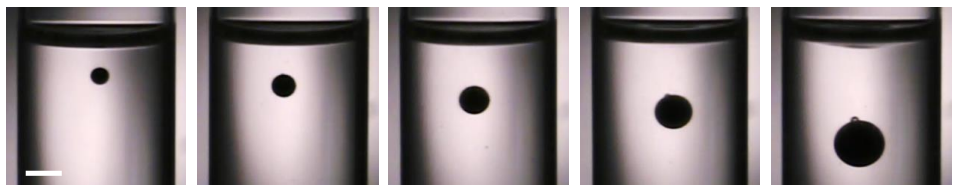

**Figure 2**. Static Stokes experiment showing the depths of submersion of Silicon Nitride Ceramic balls of diameters 2.4mm, 3.2mm, 4mm, 4.8mm and 6.35mm respectively in a 3.1% PAM hydrogel. The white scale bar represents 5mm.

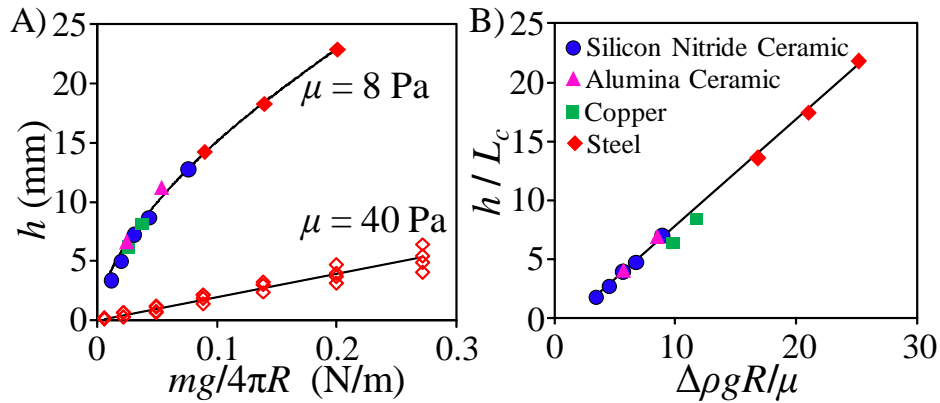

**Figure 3**. (A) The experimental data from the static immersion experiments are analyzed here by plotting the depth of submersion (*h*) against $mg/4\pi R$. Here, *m* denotes the effective mass of the spheres after correcting for buoyancy inside the gel, *g* is the gravitational acceleration and *R* is the radius of the sphere. The closed symbols represent the data obtained by performing the elastic Stokes experiment in the lower modulus (~ 8 Pa) gel whereas the open symbols represent the previously reported data[28] obtained with a higher modulus (40 Pa) gel. (B) The data for the lower modulus (8 Pa) gel are re-scaled by dividing *h* with the capillary or Laplace length ( $L_c = \sqrt{\gamma/\Delta\rho g}$ ) and plotting it against $\Delta\rho gR/\mu$. Here, $\Delta\rho$ is the difference between the density of the spheres and water, $\mu$ is the shear modulus of the gel (~8 Pa) and $\gamma$ (~73mN/m) is its surface tension.[28]

Although a power law equation ($h$ (m)=0.06($mg/4\pi R$)$^{0.6}$, *mg* is in Newtons) describes the data well, the equation is dimensionally incomplete. Close observation revealed a clear difference in the deformation in these two gels as the balls were released upon them. While for the higher modulus gel, the ball penetrates to a considerable distance inside the gel, its surface wraps around the ball incompletely, whereas the lower modulus gel completely wraps the ball and folds above it. The low value of the shear modulus (~ 8 Pa), coupled with the work of cohesion ($2\gamma$) of the gel as 144 mN/m give rise to an elasto-adhesive length ($2\gamma/\mu$) of about 20 mm, from which it is inferred that the surface tension driven auto-wetting can support a substantially long line connecting the sphere and the free surface of the gel. For this gel, excellent linear collapse of all the data can be obtained if *h* divided by the Capillary length $(\gamma/\Delta\rho g)^{0.5}$ is plotted against $\Delta\rho gR/\mu$. This linear variation of $h/(\gamma/\Delta\rho g)^{1/2}$ with $\Delta\rho gR/\mu$ suggests that *h* follows the geometric mean of two length scales: the elastocapillary length ($\gamma/\mu$) and the elastic Stokes length $\Delta\rho g R^2/\mu$. Thus, the surface tension may play some role in the depth of the submersion of the ball in the lower modulus gel, which may be owing to the large elastocapillary length of the system. However, a full resolution of the issue can only be made with an adequate non-linear (large deformation) elastic analysis of the problem, which is reserved for the future.

**3.3 Long Range Attraction Between Spheres Suspended in the Gel:**

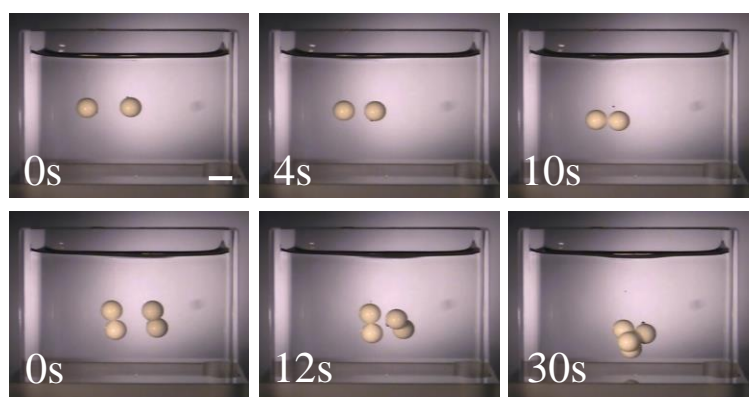

**Figure 4**. The video micrographs (upper panel) illustrate the long range attraction between two ceramic spheres (4.8 mm diameter) submerged inside a soft PAM hydrogel. The micrographs of the lower panel capture the events following the immersion of a glued dimer of similar balls inside the gel. The dimers orient (0s to 12s) as they descend inside the gel and approach each other. Finally (30s), they form a close packed structure. The white scale bar represents 5mm.( see also Movies 1 and 2 in Supporting Information (SI)).

Upon the release of a second ball at a moderate distance away from the first one, similar sequences of the above events lead to the development of tensile stress parallel to the free surface that too penetrates to a considerable distance inside the gel. However, as the tensile stress in between the two spheres can relax (figures 5 B-C), a net attraction ensues between the spheres (figure 4). As they approach each other, both the spheres sink down further in the gel till they reach a final equilibrium position. In the final state, the axis of the dimer rotates somewhat while the center of mass remaining unaffected. This observation clearly illustrates that the elastic energy corresponding to the strain parallel to the surface of the gel is being released with the concomitant decrease in the gravitational potential energy (with some increase of the elastic energy due to the deformation of the gel perpendicular to its surface). The process is by no means entirely passive as evidenced by the slow clockwise and the corresponding counter-clockwise rotations (see Movie 1, SI) of the two spheres that suggests a relative motion between gel and the spheres, i.e. material is being squeezed out of the space between the spheres.

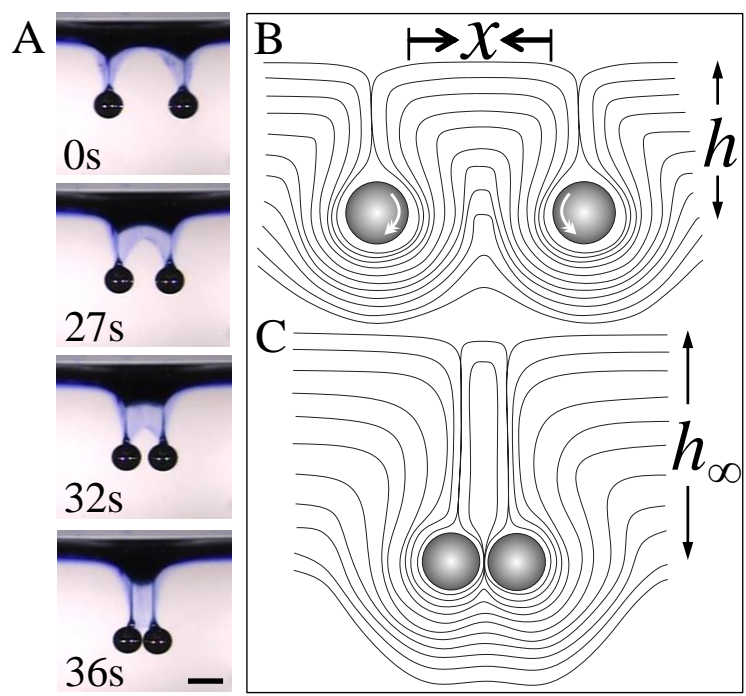

**Figure 5**. (A) The video micrographs show the attraction of two steel spheres of diameter 4mm in a chemically cross-linked gel. This experiment was performed after depositing ink on the surface of the gel with a fine needle. When a sphere is released into the gel, line formed from the folding of the surface of the gel above the ball is clearly highlighted by the intense color of the ink. Some ink is also observed around the line, which reveals that the gel in the intervening space is squeezed out as the spheres attract each other. The black scale bar represents 5mm. (see Movie 3, SI). The sequence of events in the videographs is shown schematically in (B-C). Here, $x$ denotes the distance between the balls, $h$ is the height of submersion of the balls before contact and $h_\infty$ is the height after contact.

The sequences involving the attraction and the squeeze out of the gel through the space between two spheres can be observed if ink sports are introduced on the surface of the gel before releasing the balls over it. When the balls descend through the gel, the line arising from the folding of the gel's surface above each sphere gets intensely colored as this is where the ink concentrates, which is surrounded by the region with lighter coloration by the ink. The diffusion of the ink in the gel is relatively slow, so that the squeeze out of the gel can be clearly observed (see Movie 3, SI), in that the inked gel gradually fills up the space between the lines of contact as the unmarked gel escapes (see also the schematic of figure 5) that region. The change in the total energy (elastic+potential) of the system has to be proportional to $-mg\Delta h$ in such a process. This can be shown by writing the elastic and the potential energies of a sphere as $E = \int_0^h f_{el} dh' - mgh$, where $f_{el}$ is the force on the sphere due to elastic deformation. Setting $\partial E/\partial h = 0$ leads to the equilibrium condition $mg = f_{el}$. For an elastic deformation, $f_{el} = Ch^{'n}$ ($n>1$). Substitution of this expression in the above integral leads to the result: $E = -nmgh/(n+1)$. As the spheres descend, $x$ decreases and $E$ becomes more negative.

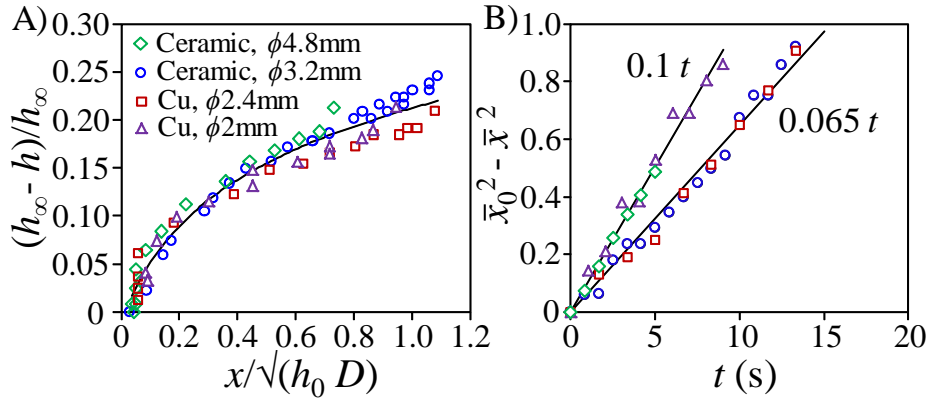

**Figure 6**. (A) This graph reveals the long range nature of the attraction of two solid spheres inside the hydrogel. Here, $D$ is the diameter of the sphere, $h_\infty$ is the depth of the two spheres after they come in full contact, $h_0$ is the initial depth of the first submerged sphere and $h$ is the average depth of the two spheres (see also the schematic of figure 1) that varies with the distance

($x$) of separation. The black curve was obtained from fitting the experimental data using Origin software, which has the following expression, $\bar{h} = 0.203 + 0.096\ln(\bar{x} + 0.104)$, where $\bar{h} = (h_\infty - h)/h_\infty$ and $\bar{x} = x/\sqrt{h_0 D}$ (B) The squared distance of separation varies linearly with time with correlation coefficients better than 98%. The symbols are same as in figure (A). $\bar{x}_0$ and $\bar{x}$ are the scaled distances of separation at times $t=0$ and $t$, respectively.

We should be able to extract a first order estimate of how attractive force varies with the distance of separation of two spheres from the variation of $-mg\Delta h$ as a function of the distance of separation. In order to make such an analysis, we measured $h$ as a function of $x$ for two identical spheres but with materials having different radii and densities. The videos of the experiments involving the attraction of the spheres were decomposed into images in VirtualDub followed by the analysis of those images to obtain the distance of separation between the spheres using the SpotTracker plugin[35] in ImageJ. Although the data obtained from each experiment could be analyzed independently to obtain the distance dependent law of attractive force, it is more convenient to process the data in terms of non-dimensional variables. In that spirit, the depth of the submersion of the ball was divided by the maximum descent of the adhered spheres [$\bar{h} = (h_\infty - h)/h_\infty$], and the distance of separation ($x$) was normalized by dividing it with $\sqrt{Dh_0}$, where $D$ is the diameter of the sphere and $h_o$ is the initial depth of descent of the first sphere. For a linear elastic system the scale $\sqrt{Dh_0}$ has the same meaning[36] as $\sqrt{mg/\mu}$. An equivalent length can also be extracted for a non-linear elastic gel as well. The experimental data plotted as above cluster around a single curve and can be fitted with a logarithmic function, i.e. $U(\bar{x}) \sim -\ln(\bar{x} + c)$, which is an asymptotic form of a more realistic modified Bessel function ( $K_0(\bar{x} + c)$) of the second kind so that $U(\bar{x})$ saturates at large separation distances. The spheres do not interact significantly at large distances, e.g. $x > 1.5$ cm. Here c is a small curve fitting constant that prevents the divergence of the interaction energy at $\bar{x} = 0$. In real situation, the cut-off can be provided by a short range repulsive force between the spheres. The force of attraction obtained from the derivative of $U(\bar{x})$ therefore follows $F(\bar{x}) \sim 1/(\bar{x}+c)$ in conformity with the observation that the interaction is rather long range. This inverse linear distance law ($F(\bar{x}) \sim 1/\bar{x}$) in conjunction with a linear kinematic friction law ($F_{drag} \sim d\bar{x}/dt$) suggest that the square of the distance of separation would decrease linearly with time, which is also observed experimentally (figure 6B).

### 3.4 Motion of a Sphere in a Thickness Graded Gel

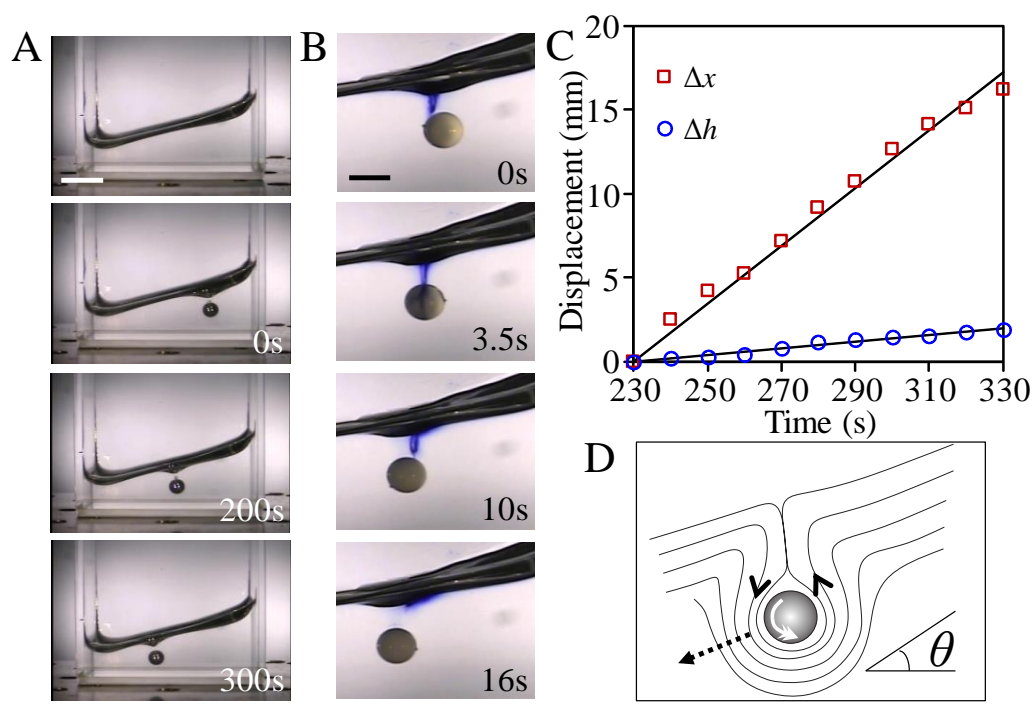

**Figure 7**. The video micrographs in (A) capture the motion of a steel ball (4 mm diameter) on a thickness graded gel, the surface of which is inclined by 14° from the horizontal plane. The gradient is constant and steepest at the central portion of the cell where both the horizontal ($x$) and the vertical ($h$) displacements of the ball increase linearly with time (C). The white scale bar represents 10mm. The micrographs in (B) capture the motion of a ceramic sphere (diameter 4.8mm) on a thickness graded gel that had ink marks. As the ball rolls down the gradient, the ink is pulled from the surface, rolls over the ball and finally returns to the surface. The black scale bar represents 5mm. (see Movie 4, SI). The schematic of the ball rolling down the graded gel is shown in (D).

The above discussions suggest that two identical spheres approach each other inside a gel following an inverse square law, and with a friction that is linear with the approach velocity. The kinematics should be contributed by such factors as the bulk dissipation in the gel as it folds and relaxes, the relative sliding of the hydrogel surfaces within the contact region, and the relative motion of the sphere inside the capsule it is in.

Motion due to the stored elastic strain energy has been observed previously in a few different settings. It was observed with an elastomeric cylinder[37] on a substrate in which the adhesion energy increased with the time of contact. As the cylinder rolled over the support, new contact was made at the advancing edge whereas the contact was broken at the trailing edge. As the receding edge of the contact reached the discontinuity of the old and of the old and new contact, the stored elastic energy caused the cylinder to roll rapidly on the substrate. Hore et al.[38] observed that differential swelling of an elastomeric rod creates the force needed to propel it uphill even carrying a load that is larger than the cylinder itself. An elastic energy gradient has also been found to be responsible for the motion[39] of defects at the interface. Style et al.[40] used a thickness graded silicone elastomer to study the phenomenon of durotaxis, in which the gradient of elastic strain energy induced a liquid drop to move on its surface. We show here that motion of an object can also be induced by the continuous folding and relaxation of the surface of a thickness graded gel. As a sphere submerges inside such an asymmetric gel, it experiences a gradient of (potential and elastic strain) energy and thus moves from the thicker towards the thinner part of the gel. This is a kind of rolling, not above a surface, but below it. If ink dots are introduced on the surface of the gel, it can be seen to be drawn inside the gel as the sphere moves that, in turn, returns to the gel's surface when the sphere passes by (see Movie 4, SI) it. Within the observation window, where the gel has a constant gradient of thickness, both the descent and the translation of the sphere in the gel increase linearly with time. Since the driving force is equal to the frictional drag force at steady state, we anticipate that $\xi(dx/dt) \sim mg(dh/dx)$, or $\xi \sim mg\dot{h}/\dot{x}^2$, where the dot indicates a derivative with respect to time. Using the data shown in figure 7c we estimate the value of $\xi$ to be 1.7 Ns/m. A comparable value of $\xi$ is also needed to explain the dynamics of attraction of two spheres as summarized in figure 6B. The solvent viscosity induced friction coefficient[41] ($\sim 8\pi\eta R$) is many orders of magnitude smaller than the above values. By considering a lubrication friction, an extremely thin water lubricating film ($\ell$ ~1nm) would be needed so that the friction coefficient $\sim \eta R^2/\ell$ is comparable to 1.7 Ns/m. These unrealistic estimates suggest that the viscous friction related to the rotation of sphere inside the capsule may not be the rate limiting factor in these experiments. The process is likely related to the viscoelastic deformation and the relaxation within the volume of the gel that is much larger than that of the sphere.

## 3.5 Interactions Between Dimers and Clusters

The long range interaction coupled with the fact that an object can freely rotate inside the capsule created by the surrounding gel lead to several interesting scenarios prevailing in complex geometries. For example, if a dimer made of two glued balls is released inside the gel that already had the dimer formed by the contact of first two spheres, they recognize each other via long range interactions and sample the most stable energetic state by orienting their axes well before coming into an intimate contact (see Movie 2, SI). The interaction between small and large spheres is equally interesting in that the smaller spheres of one kind (e.g. copper) released on the surface of the gel gets pulled into contact by a pre-existing sphere of another kind (e.g. a ceramic ball) of a larger diameter. The process continues with the sequential release of small spheres that lead to the formation of interesting patterns some of which are shown in figures 8 and 9 (see also Movies 5 and 6 in SI).

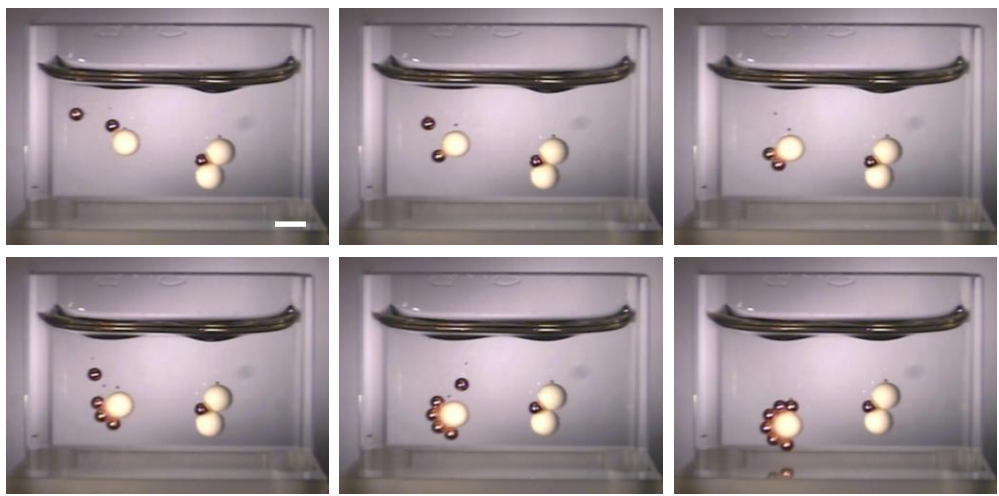

**Figure 8**. These video-micrographs capture the events leading to the formation of a semi-circular ring caused by the attraction between the copper spheres (diameter ~2.4mm) and a pre-existing ceramic sphere (4.8mm diameter) inside the gel. The copper spheres were added sequentially in the gel. The white scale bar represents 5mm. (Movie 5, SI)

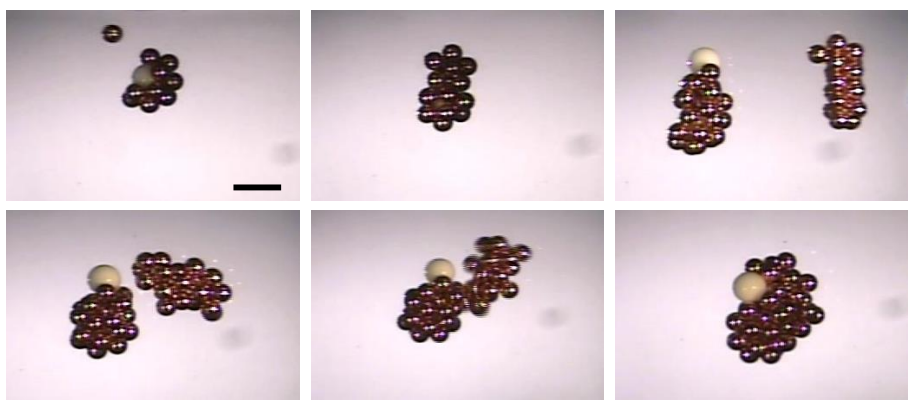

**Figure 9.** The micrographs in the upper panel show the growth of clusters on sequential addition of copper spheres (diameter 2mm) into the gel containing a ceramic sphere (diameter 3.2mm). The growing cluster eventually engulfs the ceramic sphere. When another ceramic sphere is added, it is attracted by the cluster as well. The copper spheres on their own exhibit a structure comprising of parallel columns. When the clusters are large enough (lower panel), they even attract and move towards each other. When the clusters coalesce, further re-organization of the spheres occurs that lead to a close packed state. In the online video, the abrupt stages of the re-organization of the spheres can be seen. These are reminiscent of elastic instabilities and/or plastic events. The black scale bar represents 5mm. (Movie 6, SI)

## 4. CONCLUSIONS AND OUTLOOK

The surface folding induced attraction and the motion of the particles in the soft gel are the results of the co-operative interaction of all three forces: elastic, surface tension and gravity. What we discovered here is essentially the formation of singular line defects that interact via a long range elastic field and even move towards each other thereby giving rise to an effective interaction between suspended objects. Analyses of the interactions of two spheres suggest that the attractive force follows an inverse linear distance law. While some attraction between particles is observed via elasto-capillary force even with an incomplete wrapping of the gel around the particles, the effect is more dramatic when complete wrapping occurs. This subject of particles interacting just via elasto-capillary force with the minimal intervention of gravity is itself a phenomenon worthy of study. While small particles can be considerably embedded[42] in an elastic matrix even without the effect of gravity, an interesting situation would arise with a high modulus gel or a rubber if it is compressed biaxially close to Biot's instability[43]. Since the spheres deposited on the surface of the gel or the rubber would then hardly experience any elastic resistance, we expect the conventional capillary force mediated interaction to prevail much like the case with liquids.

This work illustrates that by intercepting various material scales contributed by the elasticity, capillarity and gravity it is possible to tune in a new type of interaction that would, otherwise, not exist without the co-operative effects of its parts. Although the length scale (millimeter) explored in the current study has its own place in the repertoire of self-assembling systems[44], this philosophy of self-assembly with a tunable interaction may be extended to microscopic size objects with even softer gels and by replacing gravity with an electrical or a magnetic force, or, perhaps, even subjecting the gel to a pre-determined mechanical deformation. There is also the possibility to introduce defects in the form of localized stress concentrations, which may allow assembly of micron size particles.

In this paper, we approached the problem of elasto-capillary mediated particle interactions from the perspective of the solid like deformation of a soft matter. There, however, exists a body of beautiful works related to the interaction and aggregation of particles[45] in viscoelastic liquids for which the pre-existence of the motion of the particles is crucial. In the study reported here, interaction begins with the particles in the quiescent state. We hope to study the common features of these two manifestations of elastic correlations in an elastic gel and in a viscoelastic liquid in future.

Finally, we feel that a detailed study involving a non-linear field theory is critically needed for further expositions of the types of interactions reported here. This is, however, not a simple proposition as the magnitude of the deformations encountered here may go beyond the scope of the existing large strain neo-Hookean models. There are also issues regarding the numerical solutions of the field equations for very large deformations to be resolved. Hopefully, new theoretical and experimental studies will be inspired by the current report. Based on various observations described here, we are, however, hopeful that some simple scaling models might even emerge from rigorous treatments.

**ACKNOWLDGEMENTS**

We thank Professors Anand Jagota, Herbert Hui, Animangsu Ghatak, Hugo Caram and Ashutosh Sharma for insightful comments and discussions.

**Supporting Information Available:**

PS: All the movies are in real time. They should be viewed in full screens.

Movie 1 (Monomer Attraction)

Two alumina ceramic balls (diameter 4.8mm), are released into the PAM gel one after the other, in a plane parallel to the direction of the camera. The initial separation distance between the two balls is around 5.5mm. The two balls get attracted towards each other, each one demonstrating clockwise and counter-clockwise rotations respectively, while descending into the gel. After they make intimate contact, the physically bonded dimer orients itself to reach equilibrium.

Movie 2 (Dimer Attraction)

A glued dimer (made by joining two alumina ceramic spheres of diameter 4.8mm with super glue) is released at a distance of about 1cm away from the physically bonded dimer formed by two alumina ceramic spheres of diameter 4.8mm (in Movie 1). They twist and rotate as they come close to each other while slowly descending further into the gel to reach a final equilibrium position.

Movie 3 (Attraction of steel balls observed with ink spots on the gel)

The gel surface in this experiment was marked with spots of ink before the spheres were released into the gel. On releasing two identical steel balls (diameter 4mm) into the gel, one after another, they start moving towards each other by squeezing out some gel from the intervening space.

Movie 4 (Motion of a ceramic ball on a thickness graded gel)

A ceramic sphere (diameter 4.8mm) rolls down the thickness graded gel spotted with ink.

Movie 5 (Ring formation)

The movie shows how copper balls (diameter 2.4mm) get pulled towards a preexisting alumina ceramic ball (diameter 4.8mm) to form a semi-circular ring around it. The structure on the right was formed when a preexisting physically bonded dimer formed by two 4.8mm ceramic balls pulled towards its center a copper ball of 2.4mm diameter, when the latter was released close to it.

Movie 6 (Cluster Interaction)

A cluster (left) was formed by releasing 2mm diameter Copper balls in to the PAM gel already containing a 3.2mm diameter alumina ceramic ball. The copper balls self-assembled around the ceramic ball and pulled towards itself another 3.2mm ceramic ball when the latter was released close to it. The second cluster (right) was formed by releasing 2mm diameter Copper balls not very close to the cluster on the left, yet not too far from it. They self-assembled into a large cluster by forming several intermediate ordered column-like structures. When the two clusters are large enough, they start attracting each other by orienting and re-organizing to form a closely packed structure.

**This material is available free of charge via the Internet at http://pubs.acs.org**

For "Table of Contents Use Only"

# Surface Folding Induced Attraction and Motion of Particles in a Soft Elastic Gel: Cooperative Effects of Surface Tension, Elasticity and Gravity

Aditi Chakrabarti and Manoj K. Chaudhury[*]

Department of Chemical Engineering
Lehigh University
Bethlehem, PA 18015

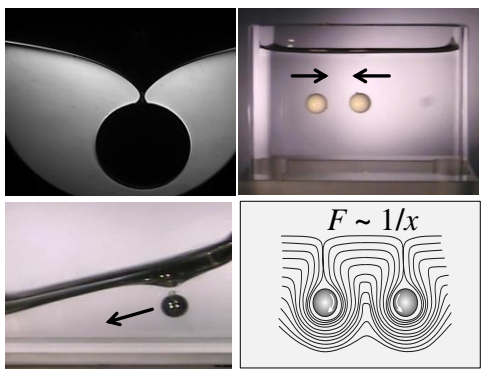